\documentclass[a4paper]{jpconf}
\usepackage{graphicx}
\usepackage{amssymb}
\begin{document}
\title{Some vortex solutions in the extended Skyrme-Faddeev model}

\author{L.~A.~Ferreira$^1$, M.~Hayasaka$^2$, J. J\"aykk\"a$^3$, N.~Sawado$^2$ and K.~Toda$^4$}
\address{$^1$ Instituto de F\'\i sica de S\~ao Carlos; IFSC/USP,Universidade de S\~ao Paulo - USP, 
Caixa Postal 369, CEP 13560-970, S\~ao Carlos-SP, Brazil}
\address{$^2$ Department of Physics, Tokyo University of Science,
 Noda, Chiba 278-8510, Japan}
\address{$^3$ Nordita, Roslagstullsbacken 23, SE-10691 Stockholm, Sweden}
\address{$^4$ Department of Mathematical Physics,
Toyama Prefectural University,
Kurokawa 5180, Imizu, Toyama, 
939-0398, Japan
}

\ead{sawado@ph.noda.tus.ac.jp~(Corresponding author)}

\begin{abstract}
Analytical and numerical vortex solutions for the extended Skyrme-Faddeev model in a $(3+1)$ dimensional Minkowski space-time are investigated. 
The extension is obtained by adding to the Lagrangian  a quartic term, which is the square of the kinetic term, and a potential which breaks 
the $SO(3)$ symmetry down to $SO(2)$.  The construction of the solutions has been done in twofold: one makes use of an axially symmetric ansatz and 
solves the resulting ODE by an analytical and a numerical way.   
The analytical vortices are obtained for special form of the potentials, and the numerical ones are computed using the 
successive over relaxation method for wider choice of the potentials. Another is based on a simulational technique named 
the simulated annealing method which is available to treat the non-axisymmetric shape of solutions. 
The crucial thing for determining the structure of vortices is {\it the type} of the potential.  
\end{abstract}

\section{Introduction}

The so-called Skyrme-Faddeev model was introduced in the seventies \cite{sf} as a
ge\-ne\-ra\-li\-za\-tion to $(3+1)$ dimensions of the $O(3)$ non-linear sigma model in
$(2+1)$ dimensions. The Skyrme term, quartic in derivatives of the field,
balances the quadratic kinetic term and according to Derrick's theorem, allows the
existence of stable solutions with non-trivial Hopf topological charges. Due to the highly
non-linear character of the model and the lack of symmetries, the first soliton solutions
were only constructed in the late nineties using numerical methods
\cite{glad,solfn,sutcliffe,hietarinta}. Since then the interest in the model has increased
considerably and it has found applications in many areas of physics due mainly to the
knotted character of the solutions \cite{babaev}. 
One of the aspects of the model that has attracted considerable
attention has been its connection with gauge theories. Faddeev and Niemi have conjectured
that it might describe the low energy limit of the pure $SU(2)$ Yang-Mills theory
\cite{fn}. They based their argument on a decomposition of the physical degrees of freedom
of the $SU(2)$ connection, proposed in the eighties by Cho \cite{chofn}, and involving a
triplet of scalar fields ${\vec n}$ taking values on the sphere $S^2$ (${\vec n}^2=1$).
Gies \cite {gies} has calculated
the Wilsonian one loop effective action for the pure $SU(2)$ Yang-Mills theory assuming
Cho's decomposition, and found that the Skyrme-Faddeev action is indeed part of it, but
additional quartic terms in the derivatives of the triplet ${\vec n}$ are unavoidable.  In
fact, the first numerical Hopf solitons were first constructed for the Skyrme-Faddeev
model modified by a quartic term \cite{glad} which is the square of the kinetic
term. However, the soliton solutions in \cite{glad} were constructed for a sector of the
theory where the signs of the coupling constants disagree with those indicated by Gies'
calculations. Therefore, it is worth investigating the model with correct sign of the 
coupling constants.

In this paper we consider an extended Skyrme-Faddeev model (ESF)
defined by the Lagrangian
\begin{eqnarray}
{\cal L} = M^2\, \partial_{\mu} {\vec n}\cdot\partial^{\mu} {\vec n}
 -\frac{1}{e^2} (\partial_{\mu}{\vec n} \wedge 
\partial_{\nu}{\vec n})^2 + \frac{\beta}{2}
\left(\partial_{\mu} {\vec n}\cdot\partial^{\mu} {\vec n}\right)^2- V(n_3)
\label{action}
\end{eqnarray}
where ${\vec n}$ is a triplet of real scalar fields taking values on the sphere $S^2$,
$n_3$ its third component, $M$ is a coupling constant with dimension of $({\rm
  length})^{-1}$, $e^2$ and $\beta$ are dimensionless coupling constants, and the
potential is a functional of the third component $n_3$ of the triplet ${\vec n}$. Note
that the potential breaks the $O(3)$ symmetry of the original Skyrme-Faddeev down to
$O(2)$, the group of rotations on the plane $n_1\, n_2$, and so eliminating two of the
three Goldstone boson degrees of freedom. In this paper the main role of potential is to
stabilize the vortex solutions.

The static energy density (${\cal H}_{{\rm static}}=-{\cal L}$) associated to (\ref{action})
is positive definite if $V>0$, $M^2>0$, $e^2>0$ and $\beta<0$. That is the sector explored
in \cite{glad} and where Hopf soliton solutions were first constructed (for $V=0$). In
addition, that is also the sector explored in \cite{Sawado:2005wa} but with additional
terms involving second derivatives of the ${\vec n}$ field, and where Hopf soliton were
also constructed. The static energy density of (\ref{action}) is also positive definite for
$V>0$ if
\begin{eqnarray}
M^2>0\,; \qquad  e^2<0\, ; \qquad \beta <0 \, ; \qquad  \beta\, e^2\geq 1
\label{nicesector}
\end{eqnarray}
That is the sector that agrees with the signature of the terms in the one loop effective
action calculated in \cite{gies} and it is the sector that we will consider in this
paper. Static Hopf solitons were constructed in \cite{sawadohopfions} for the
sector (\ref{nicesector}) (with $V=0$) and their quantum excitations, including comparison
with glueball spectrum, were considered in \cite{quantumhopfions}.  An interesting feature
of the Hopf solitons constructed in \cite{sawadohopfions} is that they shrink in size and
then disappear as $\beta\,e^2\rightarrow 1$, which is exactly the point where the exact 
vortex solution exist~\cite{vortexlaf}. 

The aim of the present paper is to investigate if vortex solutions for the model
(\ref{action}) continue to exist when the condition $\beta\, e^2=1$ is relaxed, and so if
they co-exist with the Hopf solitons of \cite{sawadohopfions}. 
In order to stabilize the solution, we shall introduce
the types of potential
\begin{eqnarray}
  V(n_3)= \frac{\mu^2}{2}v^a_b, ~~~~v^a_b\equiv (1+n_3)^{a}(1-n_3)^{b}
  \label{potdef}
\end{eqnarray}
where $a+b=$ {\it non-zero integer}, and $\mu$ is a real coupling constant.
A special choice of the parameters $a,b$ we have holomorphic solutions 
of the model while for the other case we still have numerical solutions. 

In this paper, first we discuss the integrable holomorphic solutions of the model and 
next we shall perform the numerical stuff. The numerical simulations are done by 
twofold: one is by solving a differential equation which is accomplished in terms of 
{\it the standard successive over relaxation}. Another is based on energy 
minimization scheme called {\it the simulated annealing}. Especially the latter analysis
demonstrates the detailed behavior of the symmetry breaking of the solutions by the 
change of the structure of the potential.  
  
\section{The integrable sector of the model}

The first exact vortex solutions for the theory (\ref{action}) were constructed in
\cite{vortexlaf} for the case where the potential vanishes, and by exploring the
integrability properties of a submodel of (\ref{action}).  In order to describe those exact
vortex solutions it is better to perform the stereographic projection of the target space
$S^2$ onto the plane parameterized by the complex scalar field $u$ and related to ${\vec n}$ 
by
\begin{eqnarray}
{\vec n} = (u+u^*,-i(u-u^*),u^2 -1)/(1+u^2)
\label{udef}
\end{eqnarray}
It was shown in \cite{vortexlaf} that the field configurations of the form 
\begin{eqnarray}
u\equiv u(z,y),~~u^*\equiv u^*( z^*,y),~~{\rm for}~~
\beta e^2=1,~~~~V=0 \label{exactclass}
\end{eqnarray}
are exact solutions of (\ref{action}), where $z=x^1+i\varepsilon_1x^2$ and
$y=x^3-\varepsilon_2x^0$, with $\varepsilon_a=\pm 1$, $a=1,2$, and $x^{\mu}$,
$\mu=0,1,2,3$, are the Cartesian coordinates of the Minkowski space-time. 
The simplest solution is of the form $u=z^n e^{i\,k\,y}$, with $n$ integer, 
and it corresponds to a vortex parallel to the $x^3$-axis and with waves 
traveling along it with the speed of light. 

In terms of the complex scalar field $u$ introduced in \ref{udef} the Lagrangian
\ref{action} becomes
\begin{eqnarray}
{\cal L}=
4 M^2\frac{\partial_{\mu}u \partial^{\mu}u^*}{(1+u^2)^2} + 
\frac{8}{e^2}\left[ 
\frac{(\partial_{\mu}u)^2(\partial_{\nu}u^*)^2}{(1+u^2)^4}+
(\beta e^2-1)\frac{(\partial_{\mu}u\;\partial^{\mu}u^*)^2}{(1+u^2)^4}
\right]- V(\mid u\mid^2)
\label{actionu}
\end{eqnarray}
where we have used the fact that $n_3$ is a functional of $\mid u\mid^2$ only, and so is
the potential. The Euler-Lagrange equations following from (\ref{actionu}), or (\ref{action}),
reads
\begin{eqnarray}
(1+|u|^2)\partial^{\mu}{\cal K}_{\mu}-2u^{*}{\cal K}_{\mu}
\partial^{\mu} u=-\frac{u}{4}(1+|u|^2)^3\,V^{\prime}
\label{eqmot}
\end{eqnarray}
where $V^{\prime}=\frac{\partial V}{\partial u^2}$, and 
\begin{eqnarray}
{\cal K}_{\mu}\equiv M^2 \partial_{\mu}u 
+\frac{4}{e^2}\,\frac{ 
\left[(\partial_{\nu}u\partial^{\nu} u)
\partial_{\mu}u^{*}+(\beta\,e^2-1)(\partial_{\nu}u \partial^{\nu}u^{*})
\partial_{\mu} u\right]}{(1+u^2)^2}
\label{kdef}
\end{eqnarray}
We point out that the theory (\ref{actionu}) possesses an integrable sector defined by the
condition
\begin{eqnarray}
(\partial_{\mu}u)^2=0
\label{eikonal}
\end{eqnarray}
Such condition was first discovered in the context of the $CP^1$ model using the
generalized zero curvature condition for integrable theories in any dimension \cite{afs1},
and then applied to many models with target space being the sphere $S^2$, or $CP^1$ (see
\cite{afs2} for a review). It leads to an infinite number of local conserved
currents. Indeed, (\ref{eikonal}) together with the equations of motion (\ref{eqmot}) imply the
conservation of the infinity of currents given by
\begin{eqnarray}
J_{\mu}^G\equiv {\cal K}_{\mu} \frac{\delta G}{\delta u}-{\cal K}_{\mu}^* \frac{\delta G}{\delta u^*} 
\label{infinitecurr}
\end{eqnarray}
where $G$ is any functional of $\mid u\mid^2$ only. For the case where the potential
vanishes, the set of conserved currents is considerably enlarged since $G$ can be an
arbitrary functional of $u$ and $u^*$, but not of their derivatives. If in addition to the
condition (\ref{eikonal}) one takes $V=0$ and $\beta\,e^2=1$, then the equations of motion
reduce to $\partial^2u=0$. It is in that integrable sector that the solutions
(\ref{exactclass}) lie, and were studied in \cite{vortexlaf}. 

It is interesting to note that (\ref{eqmot}) with a special choice of the potential
\begin{eqnarray}
V(n_3)=\frac{\mu^2}{2}(1+n_3)^{2-2/n}(1-n_3)^{2+2/n}
\label{potana}
\end{eqnarray}
have an analytical, holomorphic solution for each topological charge as
\begin{eqnarray}
u(\rho,\varphi,z,\tau)=\Bigl(\frac{\rho}{a}\Bigr)^ne^{i[\epsilon n\varphi+k(z+\tau)]}
\label{holomorphic}
\end{eqnarray}
where $\epsilon=\pm 1$ and $a$ describes a scale of the solution.
Here we used dimensionless polar coordinates $(\rho,\varphi,z,\tau)$ defined by
\begin{eqnarray}
x^0=c t= r_0 \tau,~~
x^1=r_0 \rho\, \cos \varphi,~~
x^2=r_0 \rho\, \sin \varphi,~~
x^3= r_0 z
\label{coord}
\end{eqnarray}
and where we have introduced   a length scale $r_0$ given by 
\begin{eqnarray}
r_0^2=-\frac{4}{M^2\,e^2}.
\label{r0def}
\end{eqnarray}
Substituting (\ref{holomorphic}) into (\ref{eqmot}) the $a$ can be determined such as
\begin{eqnarray}
a=\mid n\mid\left[\frac{M^2(\beta e^2-1)}{r_0^2\mu^2}\right]^{1/4}
=\mid n\mid\left[\frac{(-e^2)\,(\beta e^2-1)M^4}{4\,\mu^2}\right]^{1/4}
  \label{conditiona}
\end{eqnarray}
Clearly, the special solution at $\beta e^2=1$ is obtained if we take a proper limit of 
the vanishing potential, i.e.$\beta e^2\to 1$ and $\mu^2\to 0$ with $(\beta e^2-1)/\mu^2=$
constant. 

\begin{figure}[t]
\includegraphics[width=8.1cm,clip]{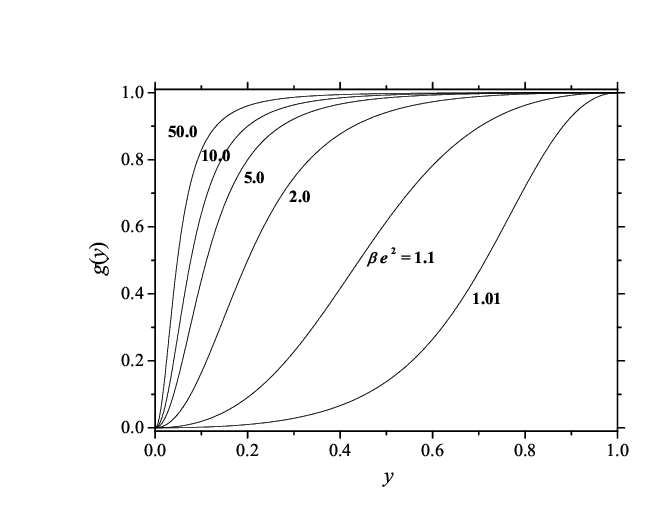}\hspace{-0.5cm}
\includegraphics[width=8cm,clip]{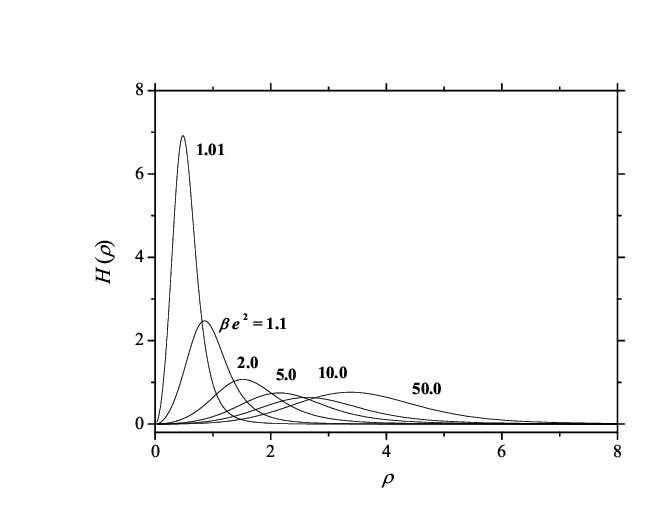}

\caption{
  \label{profiles2_v13}
  The $n=2$ profile $g(y)$ and the corresponding Hamiltonian density of the real space
  $H(\rho)$ of $k^2=0.0$ for the constant $r_0^2\mu^2/M^2=1.0$.}
\end{figure}

\begin{figure}
\hspace{3cm}\includegraphics[width=10cm,clip]{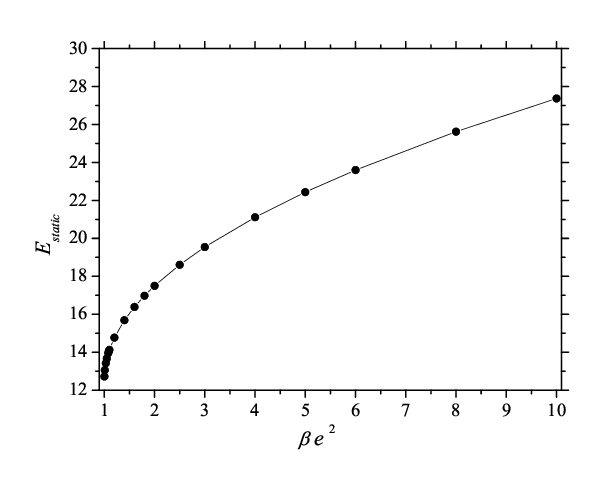}\\
\vspace{-0.0cm}
@\\
\vspace{-2cm}
@\\
\includegraphics[width=16cm,clip]{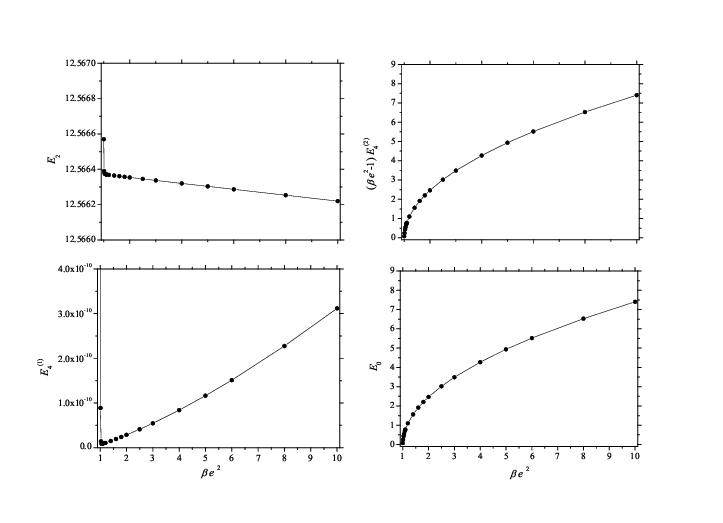}
\caption{
  \label{energy2_v13}
  The static energy and its components corresponding to the solutions of
  Fig.\ref{profiles2_v13}.}
\end{figure}

In the case with the potential, the current (\ref{infinitecurr}) is still conserved because
\begin{eqnarray}
&&\partial^\mu J_\mu=\frac{\delta^2G}{\delta u^2}\partial^\mu u{\cal K}_\mu+\frac{\delta G}{\delta u}\partial^\mu{\cal K}_\mu
+\frac{\delta^2 G}{\delta u^*\delta u}\partial^\mu u^*{\cal K}_\mu \nonumber \\
&&\hspace{1cm}-\frac{\delta^2G}{\delta u^{*2}}\partial^\mu u^*{\cal K}^*_\mu-\frac{\delta G}{\delta u^*}\partial^\mu{\cal K}^*_\mu
-\frac{\delta^2 G}{\delta u\delta u^*}\partial^\mu u^*{\cal K}^*_\mu=0
\end{eqnarray}
where we have used the reduced integrable equation 
\begin{eqnarray}
\partial^\mu{\cal K}_\mu+\frac{\mu^2}{4}(1+|u|^2)^2\frac{\partial V(|u|^2)}{\partial u^*}=0.
\end{eqnarray}

The Hamiltonian density associated to (\ref{actionu}) is not positive definite due to the
quartic terms in time derivatives. We shall arrange the Legendre transform of each term in
(\ref{actionu}) to make explicit such non positive contributions, and write the Hamiltonian
density as (see \cite{Ferreira:2009vi} for details)
\begin{eqnarray}
  {\cal H} &=& 4 M^2 \frac{\left[\mid{\dot u}\mid^2+{\vec \nabla}u\cdot
      {\vec\nabla}u^*\right]}{(1+u^2)^2} -\frac{24}{e^2} \frac{({\vec \nabla}u)^2
    ({\vec \nabla}u^*)^2}{(1+u^22)^4} \left[ \Bigl(\frac{2}{3}\Bigr)^2-F^2 \right] \nonumber \\
   &-&24 \frac{(\beta e^2-1 )}{e^2} \frac{\left[\mid{\dot u}\mid^2+ \frac{1}{3} 
       {\vec \nabla}u\cdot {\vec\nabla}u^*\right] \left[ {\vec \nabla}u\cdot
      {\vec\nabla}u^*-\mid{\dot u}\mid^2\right]}{(1+u^2)^4} +V(\mid u\mid^2)
  \label{energy}
\end{eqnarray}
where ${\dot u}$ denotes the $x^0$-derivative of $u$, and ${\vec \nabla}u$ its spatial
gradient, and where we have denoted
\begin{eqnarray}
  \frac{{\dot u}^2}{({\vec \nabla}u)^2}\equiv \frac{1}{3} + F e^{i\,\Phi}
  \label{fphidef}
\end{eqnarray}
with $F>0$ and $0\leq\Phi\leq 2\pi$, being functions of the space-time coordinates.  
The most of terms in (\ref{energy}) make positive contribution while the second term has some 
possibility to be negative.  
Note also, for static configurations apparently it is positive definite for the range of 
parameters given in (\ref{nicesector}).

\section{The integrable and the non-integable sectors: numerical analysis by the SOR}

Although in the previous section we used the polar coordinates, for the numerical study it is more convenient
to use a new radial coordinate $y$ ($0\leq y\leq 1$), defined by $\rho=\sqrt{\frac{1-y}{y}}$. We introduce the solution ansatz 
of the form
\begin{eqnarray}
u(\rho,\varphi,z,\tau)=\sqrt{\frac{1-g(y)}{g(y)}}e^{i[\epsilon n\varphi+k(z+\tau)]}
\label{solnu}
\end{eqnarray}
where the profile function $g(y)$ is defined at the period $0\leq g\leq 1$.
The equation can be written as 
\begin{eqnarray}
  &&\frac{d}{dy}\biggl[\frac{y(1-y)}{g(1-g)}g'R\biggr]
  +\Bigl(g-\frac{1}{2}\Bigr)\frac{S}{y(1-y)}\biggl\{\Omega-\biggl(\frac{y(1-y)}{g(1-g)}g'\biggr)^2\biggr\} \nonumber \\
  &&\hspace{2cm}=-\frac{1}{y^2}\frac{r_0^2\mu^2}{M^2}(1-g)^{1-\frac{2}{n}}g^{1+\frac{2}{n}}\Bigl\{4g-2\Bigl(1+\frac{1}{n}\Bigr)\Bigr\}
  \label{eqforg}
\end{eqnarray}
where the primes at this time indicate derivatives w.r.t.$y$ and where
\begin{eqnarray}
  &\Omega=n^2 \nonumber \\
  &S=1+\beta e^2g(1-g)\frac{y}{1-y}\biggl\{\Omega+\biggl(\frac{y(1-y)}{g(1-g)}g'\biggr)^2\biggr\} \nonumber \\
  &R=1+g(1-g)\frac{y}{1-y}\biggl\{(\beta e^2-2)\Omega+\beta
  e^2\biggl(\frac{y(1-y)}{g(1-g)}g'\biggr)^2\biggr\} 
\label{eqforg2}
\end{eqnarray}

\begin{figure*}
\centering
\includegraphics[width=7cm,clip]{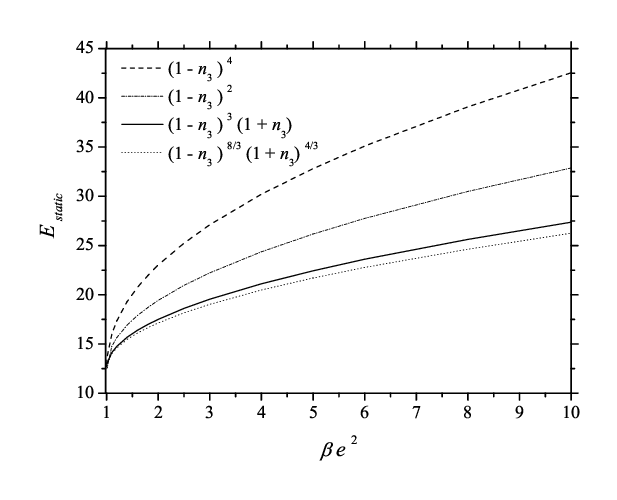}
\includegraphics[width=7cm,clip]{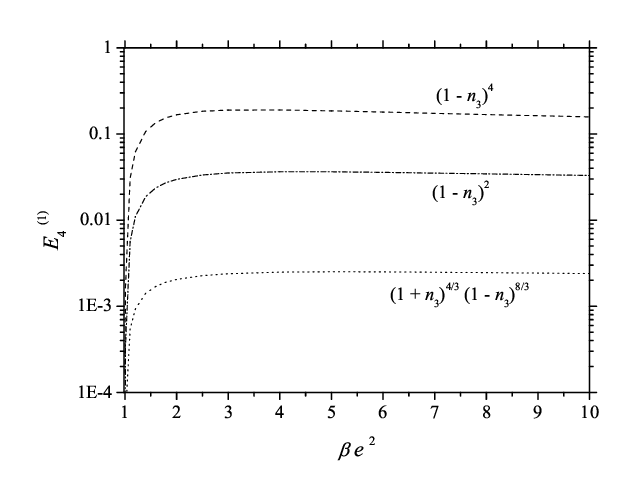}
\caption{
  \label{energy2_pots}
  The $n=2$ static energy and the component $E_4^{(1)}$ for several type of potentials
  $v^i_j$.  $k^2=0$ and $r_0^2\mu^2/M^2=1.0$.}
\end{figure*}

The energy in the unit of $4M^2$ per unit length for the time-dependent vortex can be estimated 
in terms of following four parts of integrals
of the dimensionless Hamiltonian $H:={\cal H}/4M^2$
\begin{eqnarray}
  E&=&2\pi \int^\infty_0 \rho d\rho H(\rho)=E_2+E_4^{(1)}+(\beta e^2-1)E_4^{(2)}+\frac{r_0^2\mu^2}{M^2}E_0
  \label{energy_split}
\end{eqnarray}
in which the components are defined as
\begin{eqnarray}
  &&E_2=\pi\int^1_0\frac{dy}{y(1-y)}\biggl\{2k^2\frac{1-y}{y}+n^2+\biggl(\frac{y(1-y)}{g(1-g)}g'\biggr)^2\biggr\}g(1-g) 
  \label{ecomp2}\\
  &&E_4^{(1)}=\pi\int^1_0\frac{dy}{2(1-y)^2}\biggl\{4k^2\frac{1-y}{y}+n^2-\biggl(\frac{y(1-y)}{g(1-g)}g'\biggr)^2\biggr\}\nonumber \\
  &&\hspace{3cm}\times \biggl\{n^2-\biggl(\frac{y(1-y)}{g(1-g)}g'\biggr)^2\biggr\}\bigl(g(1-g)\bigr)^2 
  \label{ecomp41}\\
  &&E_4^{(2)}=\pi\int^1_0\frac{dy}{2(1-y)^2}\biggl\{4k^2\frac{1-y}{y}+n^2+\biggl(\frac{y(1-y)}{g(1-g)}g'\biggr)^2\biggr\}\nonumber \\
  &&\hspace{3cm}\times \biggl\{n^2+\biggl(\frac{y(1-y)}{g(1-g)}g'\biggr)^2\biggr\}\bigl(g(1-g)\bigr)^2 
  \label{ecomp42}\\
  &&E_0=2\pi\int^1_0\frac{dy}{y^2} g^{2+\frac{2}{n}}(1-g)^{2-\frac{2}{n}}.
  \label{ecompp}
\end{eqnarray}
It is easy to see that the $k^2$ term in (\ref{ecomp42}) has positive contributions to the energy while in (\ref{ecomp41}) the
sign of the $k^2$ term depends on the spatial structure of the solution $g$. 
For the holomorphic solution (\ref{holomorphic}), the energy (\ref{ecomp41}) becomes zero and 
then the energy of the integrable sector keeps positive definite for all values of $k^2$.

For $n=2$, the explicit form the potential is 
\begin{eqnarray}
V_{n=2}=\frac{\mu^2}{2}(1+n_3)(1-n_3)^3\,.
\end{eqnarray}
The potential has zero at both the origin and the infinity thus it is so called a {\it new}-BS (Baby-Skyrmion) type. 
The equation (\ref{eqforg}),(\ref{eqforg2}) are solved in terms of the standard successive over relaxation scheme. 
Fig.\ref{profiles2_v13}
is the profile function and the Hamiltonian density for $n=2$.   
Fig.\ref{energy2_v13} is the energy per unit length and its components for several 
values of $\beta e^2$ and fixed $\mu^2$.  We confirmed that the value of the 
component $E_1^{(4)}$ is regarded as zero within the numerical uncertainty.
This clearly indicates that the solution satisfies the condition (\ref{eikonal}). 

Although we have obtained the analytical solutions for a special form of the potential
(\ref{potdef}), we have many options for choice of the potential.  
We can obtain many numerical solutions for the several types of the potentials.  We show
the result of $n=2$ for the potentials $v^0_2,v^0_4,v^{4/3}_{8/3}$; of course these are
not of the form of the analytical solution. 
Note that for such non-integrable solutions, the second term in (\ref{energy}) survives and has
a possibility to be negative. Here we only consider the case for $k^2=0$.
Fig.\ref{energy2_pots} presents the energies
and the component $E_4^{(1)}$ for these potentials. For $n=2$, the {\it old}-BS potentials
give higher total energy than the {\it new}-BS. This indicates that the same class of
potentials gives the similar energy and then, for $n=2$ the energy of the {\it new} type
potential $v^{4/3}_{8/3}$ is closest to the integrable sector, which is also plotted in
Fig.\ref{energy2_pots} for reference.

\begin{figure*}
\includegraphics[width=8cm,clip]{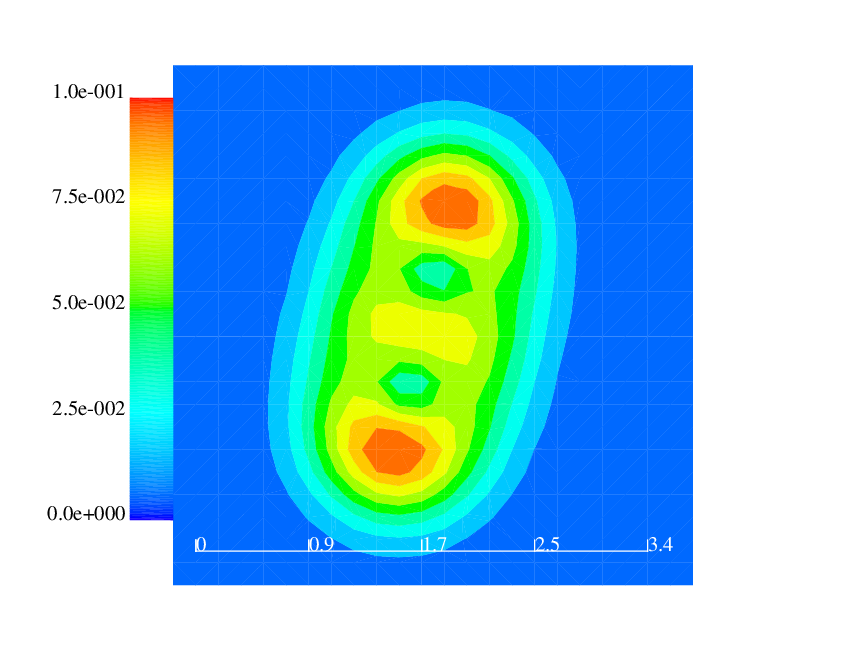}\hspace{-5mm}
\includegraphics[width=8cm,clip]{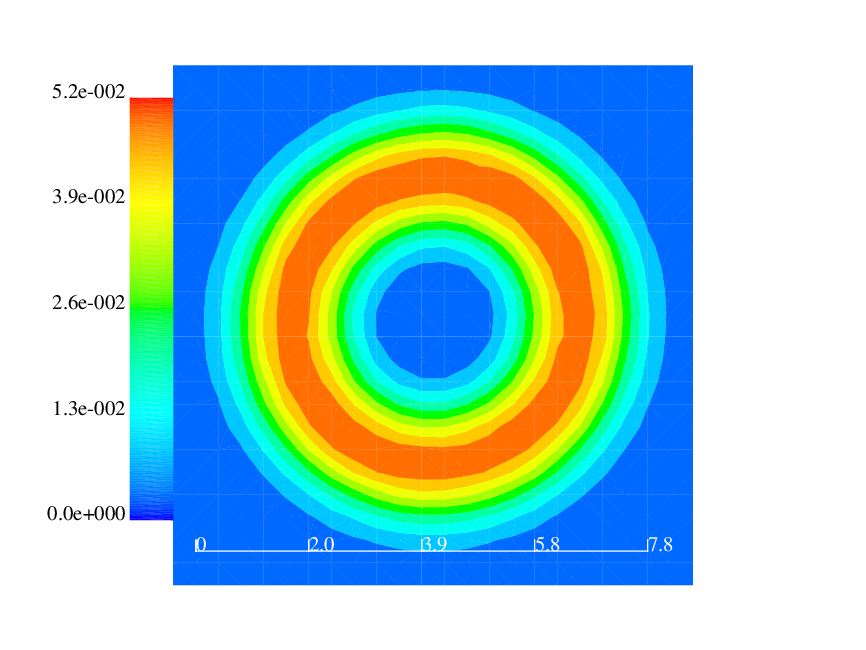}
\caption{
  \label{energy_anneal}
  The energy density per unit length of the {\it old}-BS potential $v^0_1$ ({\it left}) 
and the {\it new}-BS potential $v^1_1$ ({\it right}).}
\end{figure*}

\begin{figure*}
\includegraphics[width=4.7cm,clip]{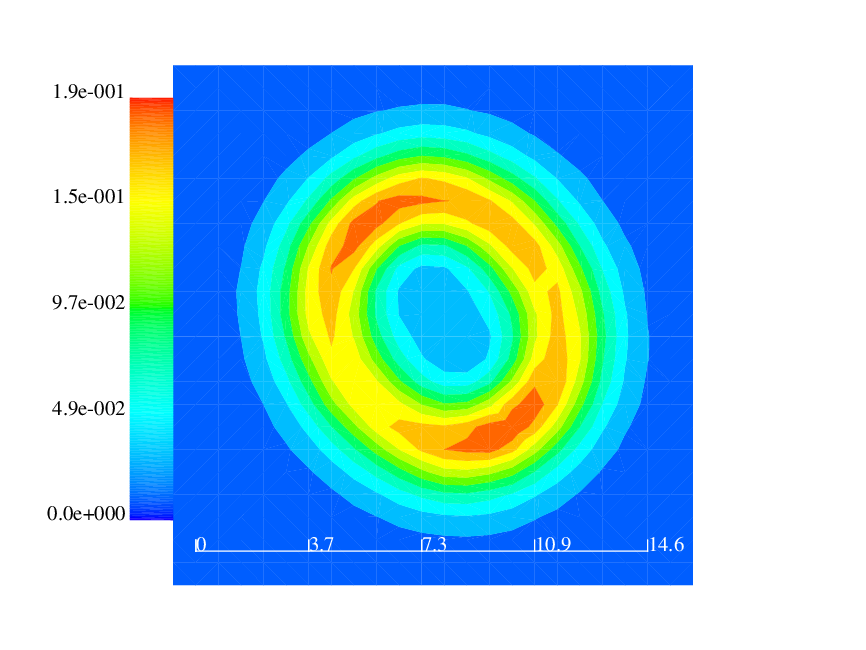}\hspace{-11mm}
\includegraphics[width=4.7cm,clip]{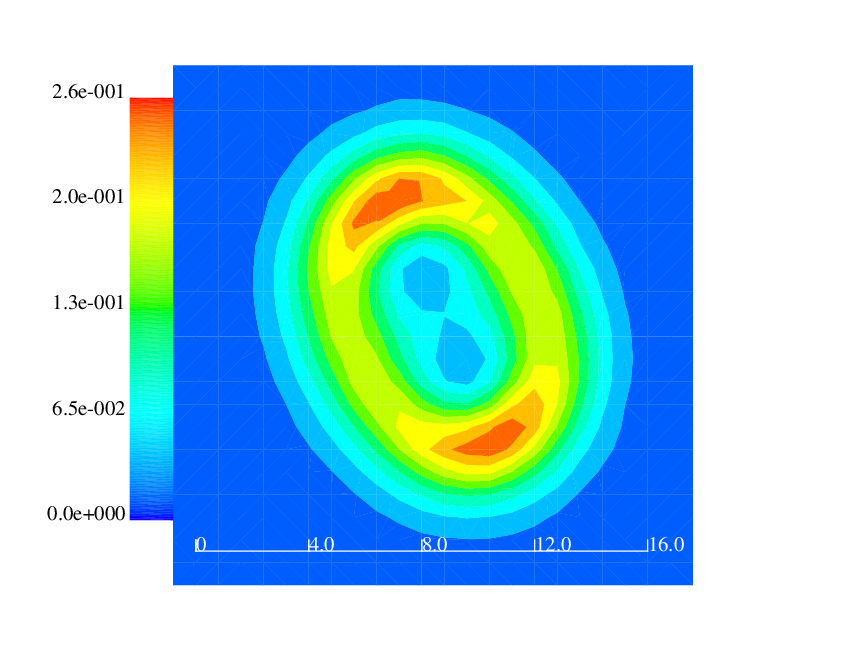}\hspace{-11mm}
\includegraphics[width=4.7cm,clip]{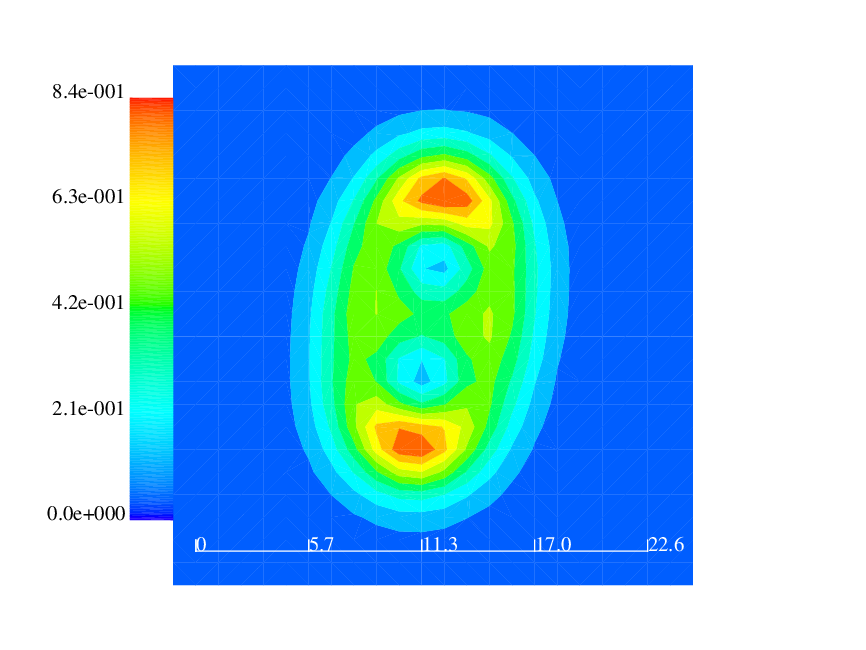}\hspace{-11mm}
\includegraphics[width=4.7cm,clip]{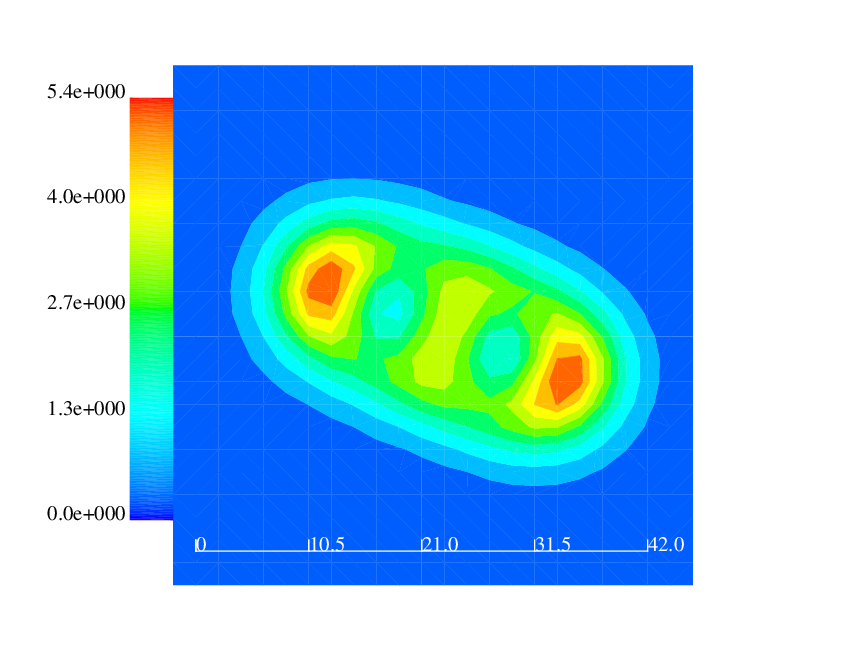}\\
$~~~~$ \hspace{0.7cm} $\beta e^2=1.01$
\hspace{1.7cm} $\beta e^2=1.1$
\hspace{1.8cm} $\beta e^2=2.0$
\hspace{1.8cm} $\beta e^2=20.0$
\caption{
  \label{energy_anneal2}
  The energy density per unit length of the {\it old}-BS potential $v^0_1$ for several $\beta e^2$.}
\end{figure*}

\section{Broken axisymmetric solutions: analysis by the SA}

In the previous sections, we have assumed that the solution is invariant under the 
$O(2)$ internal symmetry $u\to e^{i\phi}u$ as well as the transformations of the 
Poincar\'e group given by rotation on the plane $x^1x^2$ and translations in the 
directions $x^0$ and $x^3$. However, there is a possibility of relaxing some of 
these symmetries. In \cite{Hen:2007in}, the authors generalized the 
{\it old}-BS type potential as $V=\frac{\mu^2}{2}(1-n_3)^s~(0<s\leq 4)$ and saw how the solution 
deforms from the rotational symmetry on the $x^1x^2$ plane by the change of $\mu$ or $s$.

For the deformed solution, 
the straightforward generalization of the ansatz (\ref{solnu}) is
\begin{eqnarray}
u=f(\rho,\varphi)e^{i\Theta(\rho,\varphi)}e^{ik(z+\tau)}
\end{eqnarray}
or equivalently, 
\begin{eqnarray}
\vec{n}=(\sin F(\rho,\varphi)\cos [\Theta(\rho,\varphi)+k(z+\tau)],
\sin F(\rho,\varphi)\sin [\Theta(\rho,\varphi)+k(z+\tau)],\cos F(\rho,\varphi))
\label{ansatz_n}
\end{eqnarray}
where the $\vec{n}$ field in terms of $u$ is given in (\ref{udef}). Here  $\Theta (\rho,\varphi):=n\varphi+\Theta_0(\rho,\varphi)$ and $\Theta_0$ is homotopic to the 
constant map. The method is a kind of the Monte-Carlo method in which one generates 
random numbers and properly change the value of the fields $\vec{n}$ by the numbers so as to drop the energy.
However, a more sophisticated method may be applied to the problem.  
The simulated annealing method \cite{Hale:2000fk} is the application of the Metropolis algorithm 
which can successfully avoids the unwanted saddle points.
The analysis has been done by minimizing the energy per unit length  
which is obtained by integrating the hamiltonian corresponding to (\ref{action}) with the ansatz (\ref{ansatz_n}) 
into the $(\rho,\varphi)$ plane. We have confirmed that for the holomorphic solution (\ref{holomorphic})
the hamiltonian is positive definite. For the general solution, however, the energy positivity is supported only 
for a certain range of $k$. Here we present the results for $k=0$. Also, we examine the case of $n=3$ because the 
solution exhibits obvious symmetry breaking such as from axial to $\mathbb{Z}_2$-symmetry.  

\begin{figure*}
\includegraphics[width=8cm,clip]{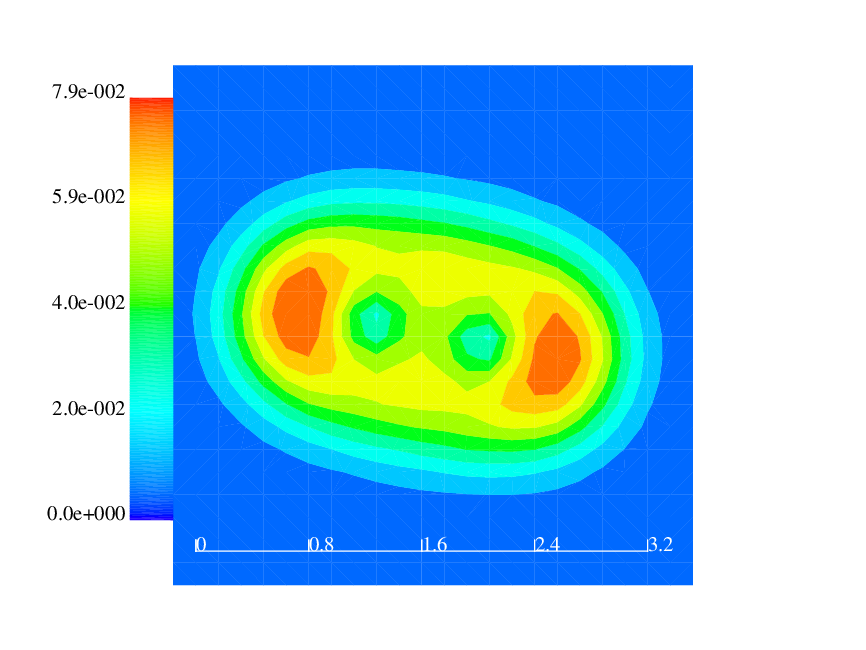}\hspace{-5mm}
\includegraphics[width=8cm,clip]{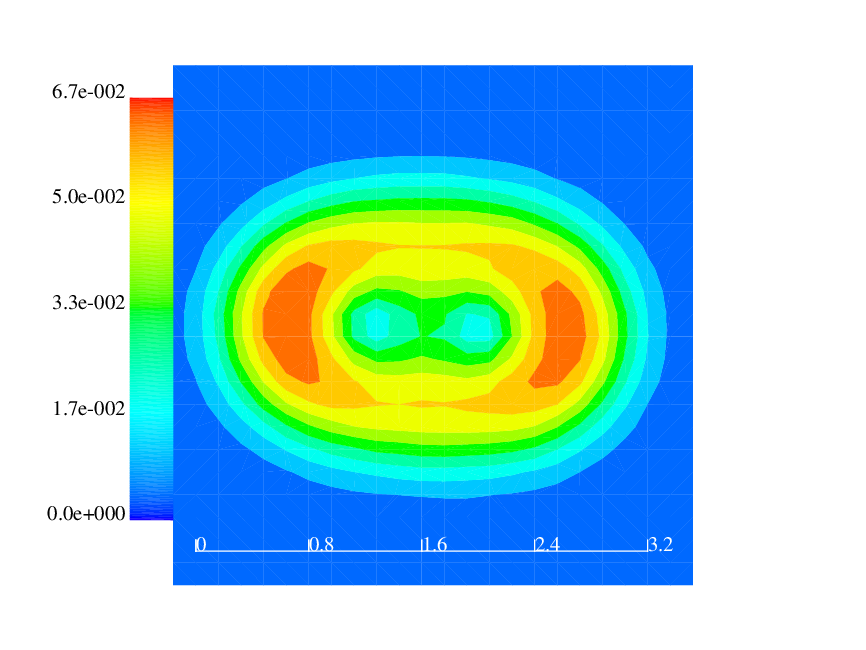}
\caption{
  \label{energy_anneal_e}
  The energy density per unit length of $V_{\epsilon}=\frac{\mu^2}{2}(1-n_3)(1+\epsilon n_3)$
for $\epsilon=0.25$ ({\it left}) and $\epsilon=0.5$ ({\it right}). The results for 
$\epsilon=0,0,1.0$ are already presented in Fig.\ref{energy_anneal}.}
\end{figure*}

\begin{figure*}
\includegraphics[width=8cm,clip]{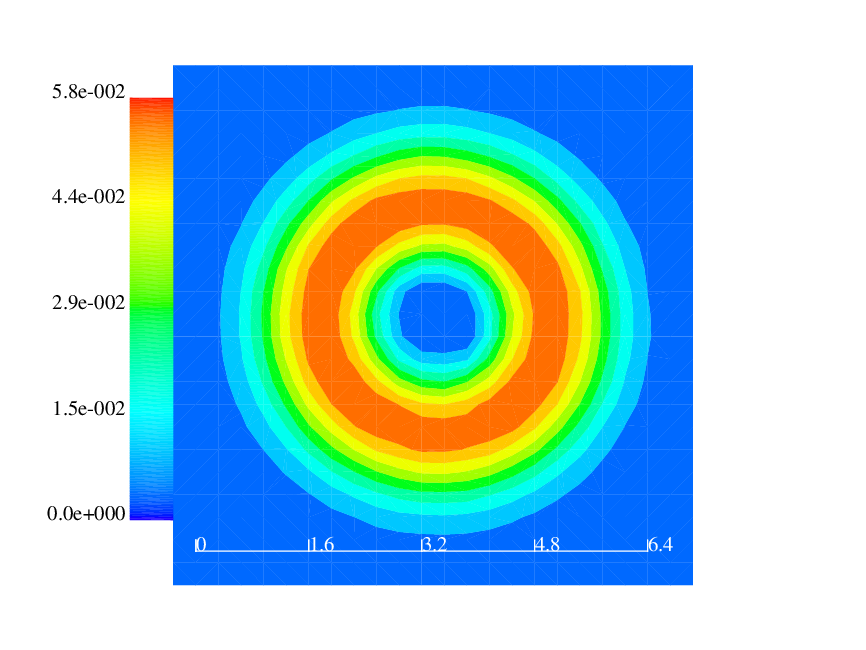}\hspace{-5mm}
\includegraphics[width=8cm,clip]{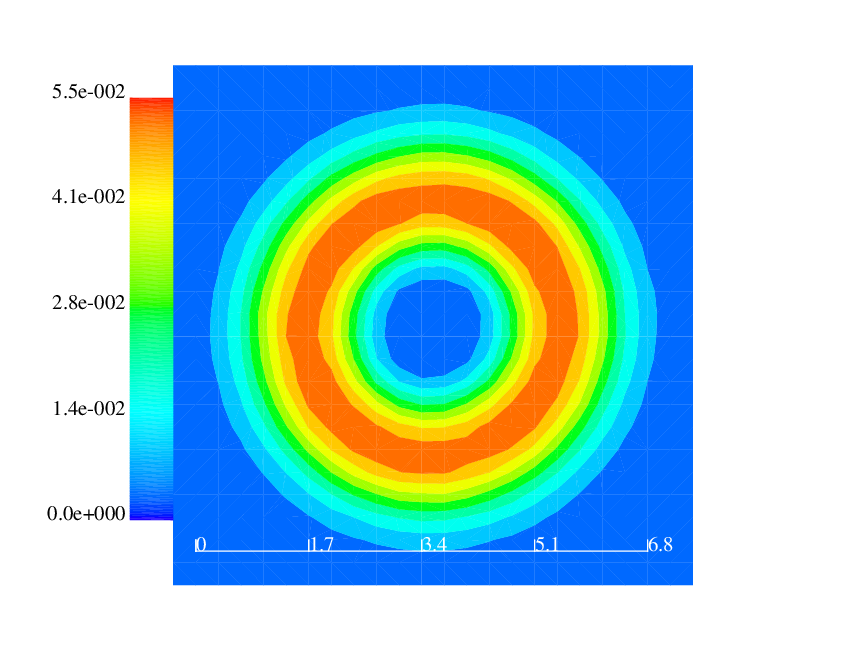}
\caption{
  \label{energy_anneal_d}
  The energy density per unit length of $V_\delta=\frac{\mu^2}{2}(1-n_3)(1+n_3)^\delta$
for $\delta=0.25$ ({\it left}) and $\delta=0.5$ ({\it right}). The results for
$\delta=0.0,1.0$ are already presented in Fig.\ref{energy_anneal}.}
\end{figure*}

We study the following several cases:
\begin{itemize}
\item[(i)]~We start with two standard cases: the {\it old}-BS potential $v^0_1$ and the {\it new}-BS
potential $v^1_1$. 
In Fig.\ref{energy_anneal}, we present the energy density per unit length. For the
{\it old}-BS potential, the solution strongly deforms from the axial symmetry while for the {\it new}-BS 
the solution keeps the symmetry. 
\item[(ii)]~
Next we shall see how the solution behaves for the change of the model parameters. 
In Fig.\ref{energy_anneal2}, we plot the energy density per unit length 
of the {\it old}-BS potential for several values of the model parameter $\beta e^2$. 
As is easily observed that 
for larger value of $\beta e^2$, the deformation is enhanced. 

\item[(iii)]~
From the results of (i),(ii), we get a new insight for the symmetry breaking. That is, 
most crucial thing for the deformation is that the potential is finite or not at the point antipodal to 
the vacuum. 
In order to investigate this criterion further, we introduce following two types of one-parameter family of potentials
\begin{eqnarray}
&&V_{\rm \epsilon}(n_3):=\frac{\mu^2}{2}(1-n_3)(1+\epsilon n_3),~~0\leq \epsilon \leq 1 \\
&&V_{\rm \delta}(n_3):=\frac{\mu^2}{2}(1-n_3)(1+n_3)^\delta,~~0\leq \delta \leq 1
\end{eqnarray} 
Note that for $\epsilon,\delta=0$ the potentials become the {\it old}-BS and $\epsilon,\delta=1$ they are
the {\it new}-BS. The results are stimulating. In the case of $V_{\rm \epsilon}$, 
the solutions always break the axial symmetry except only for $\epsilon=1.0$ (Fig.\ref{energy_anneal_e}). 
On the other hand, in the $V_{\rm \delta}$, the solutions always keep the axial symmetry except only for $\delta=0.0$ (Fig.\ref{energy_anneal_d}).
Thus we confirmed the criterion for the mechanism of the symmetry breaking: if a potential has the vacuum value at the point 
antipodal to the true vacuum, the solution always exhibits the axial symmetry, and if there is no another 
zero except for the vacuum, the solution deforms.  
\end{itemize}

We summarize our results: for $n=1$, the potential for the integrable sector is the {\it old}-BS type and 
the solution naturally has the axial symmetry. For $n\geq 2$ the potential for the integrable sector is always 
the {\it new}-BS type, and in terms of the above simulational study, the solutions always should be axially symmetric. 
As a result, our prescription of the ansatz (\ref{solnu}) is valid for all topological charges.

\noindent {\bf Acknowledgments} 
Nobuyuki Sawado expresses his gratitude to the conference organizers of ISQS-20 for kind accommodation and hospitality. 
The authors would like to thank Wojtek Zakrzewski and Pawe\l~Klimas for many useful 
discussions. 
We also acknowledge financial support from FAPESP (Brazil). Juha J\"aykk\"a appreciate 
the Flemish Science Foundation (FWO) for the support.

\end{document}